\documentclass[aps,preprint]{revtex4}%
\usepackage{amsfonts}
\usepackage{amsmath}
\usepackage{amssymb}
\usepackage{graphicx}%
\providecommand{\U}[1]{\protect\rule{.1in}{.1in}}
\begin{document}
\title{{\Large Time-dependent coherent squeezed states in a nonunitary approach}}
\author{A. S. Pereira$^{1,\ast}$}
\author{A. S. Lemos$^{2,3,\dag}$}
\affiliation{$^{1}$Instituto Federal do Par\'{a} - Concei\c{c}\~{a}o do Araguaia - PA -
Brazil }
\affiliation{$^{2}$Departamento de F\'{\i}sica, Universidade Estadual da Para\'{\i}ba -
Campina Grande - PB - Brazil.}
\affiliation{$^{3}$Departamento de F\'{\i}sica, Universidade Federal de Campina Grande -
Campina Grande - PB - Brazil}


\begin{abstract}
In this work, we have applied the integrals of motion method in a nonunitary
approach and so obtained the time-dependent displacement and squeezed
parameters of the coherent squeezed states (CSS). On its turn, CSS for
one-dimensional systems with general time-dependent quadratic Hamiltonian are
constructed. We discuss the properties of these states, in particular,
minimization of uncertainty relation and transition probabilities. As an
application, we calculate the CSS of an oscillator with a time-dependent
frequency and shown that the solution can be obtained from these well-known
Mathieu's equation.

\end{abstract}
\maketitle
\email{[$\ast$]alfaspereira@gmail.com}
\email{[$\dagger$]adiellemos@gmail.com}

\section{Introduction}

The coherent states (CS) have attracted a renewed interest in recent years. In
modern quantum physics, they play a fundamental role due to their useful
properties and the intrinsic relationship with the description of quantum
systems in a semiclassical scenario. Therefore, as a consequence, have been
applied in a wide range of studies extending from the quantum theory of
radiation \cite{Kla1968,Scu1997}, mathematical physics \cite{Kla1985}, even
quantum computation \cite{Nel2000}. There is a well-developed scheme of
constructing the CS for systems with quadratic Hamiltonians, see
\cite{Gla1963,Per1986}. In particular, the CS are defined by minimizing
uncertainty relations for some physical quantities (e.g., position and
momentum), with the same standard deviation for each of these physical
quantities. Additionally, the minimum uncertainty found for CS is identical to
that calculated from the vacuum state.

On the other hand, the CSS are quantum states for which, given specific pairs
of physical quantities, the standard deviation evaluated in one of them is
smaller than for a CS, provided that the standard deviation into another
quantity is increased. As the outcome, the CSS has been applied to improve
optical communications \cite{Yue1976}, quantum information \cite{Slu1990}, and
also are essential for the detectors of gravitational waves
\cite{Cav1981,Ni1987,Chu2014}. A scheme of constructing the CSS has been
developed through the unitary displacement and squeezing operators, which act
on the vacuum state see, e.g., \cite{Wal1983,Fis1984,Sat1985,Jan1988,Nie2000}.
See also \cite{Dod2002,Dod2003,Per2018} and the selected articles there.

The time-dependent quadratic Hamiltonian systems have attracted attention over
the years because of their usefulness in describing the dynamics of many
phenomena in quantum mechanics, quantum optics, see, e.g.,
\cite{Ber1983,Sal1989,Dod1989,Wol1990,Aga1991} and therefore offers many
applications in various fields of Physics. In the papers
\cite{Yue1976,Yeo2003,Bag2015}, the authors solved the Schr\"{o}dinger
equation with a time-dependent general quadratic Hamiltonian. In particular,
using the integral of motion method \cite{Dod1972,Mal1975,Mal1979}, some kinds
of CSS were constructed in \cite{Bag2015}. In this work, a particular set of
CSS are constructed as eigenvectors of the integrals of motion with a
corresponding complex eigenvalue.

In another way, the CSS also can be constructed using the nonunitary approach
\cite{Wun1992,Roy1992}. This method consists of introducing a nonunitary
exponential operator composed of a part of the displacement and squeezed
standard operators, which allows us to introduce the most convenient
displacement and squeezing parameters. From this procedure, we propose to use
the integral of motion method in a nonunitary approach to constructed
time-dependent CSS of a time-dependent general quadratic Hamiltonian and its
relation with the time-independent Fock-states. We must emphasize that such a
relationship is not clearly established in the unitary approach. In this
context, there are potential applications in the study of finite-level
systems, which have relation, e.g., to the problem of one- and two-qubit gates
\cite{Brem2002}, in the semiclassical theory of laser beams \cite{Nuss1973},
in optical resonance \cite{Alle1975}.

In the present paper, we are interested in investigating the properties of the
CSS for a general time-dependent quadratic Hamiltonian. Thus, this work is
organized as follows. In Sect. 2, following the integral of motion method, we
will construct the integrals of motion for this system. On its turn, this
allows us to introduce time-dependent displacement and squeezing parameters of
CSS in a nonunitary approach. Then, in Sect. 3, we will obtain the
representation for the states in terms of Fock-states and the corresponding
transition probability. In Sect. 4, we will discuss semiclassical features and
coordinate representation of the constructed CSS. Finally, in Sect. 5, as an
application of the general construction, we will consider CSS of the
time-dependent harmonic oscillator, whose solution, as we shaw see,
corresponds to that obtained from the well-known Mathieu's problem.

\section{Integrals of motion}

The time-dependent general quantum quadratic Hamiltonian system with one
degree of freedom in terms of the annihilation $\hat{a}$ and creation $\hat
{a}^{\dagger}$ operators $\left(  \left[  \hat{a},\hat{a}^{\dagger}\right]
=1\right)  $ is written as%
\begin{equation}
\hat{H}=\frac{1}{2}\hbar\left(  \alpha^{\ast}\hat{a}^{2}+\alpha\hat
{a}^{\dagger2}\right)  +\hbar\beta\hat{a}^{\dagger}\hat{a}+\hbar\gamma^{\ast
}\hat{a}+\hbar\gamma\hat{a}^{\dagger}+\hbar\delta, \label{1}%
\end{equation}
where $\alpha$, $\beta$, $\gamma$, and $\delta$ are time-dependent functions
and the signs $\dagger$ and $\ast$ denote Hermitian and complex conjugation,
respectively. From the hermiticity condition, we have that $\beta$ and
$\delta$ must be real functions.

On its turn, the quantum states $\left\vert \Psi\right\rangle $ which describe
the time evolution of the system should satisfy the Schr\"{o}dinger's equation%
\begin{align}
&  \hat{\Lambda}\left\vert \Psi\right\rangle =0,\text{ \ }\nonumber\\
&  \hat{\Lambda}=\hat{H}-i\hbar\partial_{t},\text{ \ }\partial_{t}%
=\frac{\partial}{\partial t}, \label{2}%
\end{align}
where $\hat{\Lambda}$ we call of equation operator.

Let us consider a time-dependent operator $\hat{A}=\hat{A}\left(  t\right)  $:%
\begin{align}
&  \hat{A}=f\hat{a}+g\hat{a}^{\dagger}+\varphi,\nonumber\\
&  \left[  \hat{A},\hat{A}^{\dagger}\right]  =1,\text{ \ }\left\vert
f\right\vert ^{2}-\left\vert g\right\vert ^{2}=1. \label{3}%
\end{align}
Here $f=f\left(  t\right)  $, $g=g\left(  t\right)  $ and $\varphi
=\varphi\left(  t\right)  $ are some complex function of $t$, which will be
further chosen so that the operator $\hat{A}$ to be integral of motion of the
Eq. (\ref{2}), and implies that%
\begin{equation}
\overset{\cdot}{\hat{A}}=\frac{i}{\hbar}\left[  \hat{\Lambda},\hat{A}\right]
=0, \label{4}%
\end{equation}
where, in this case, dot denotes the total derivative with respect to time $t
$.

From (\ref{3}), one can express the operator $\hat{a}$ in terms of the
integrals of motion $\hat{A}$ and $\hat{A}^{\dagger}$,%
\begin{equation}
\hat{a}=f^{\ast}\hat{A}-g\hat{A}^{\dagger}+u,\text{ \ }u=g\varphi^{\ast
}-f^{\ast}\varphi. \label{8}%
\end{equation}

Let us consider the generalized coordinate $x$ on the whole real axis and the
canonical momentum $\hat{p}=-i\hbar\partial_{x}$ in terms of the operators
$\hat{a}$ and $\hat{a}^{\dagger}$ in the form:%
\begin{align}
&  \hat{a}=\frac{1}{\sqrt{2}}\left(  \frac{\hat{x}}{l}+\frac{il}{\hbar}\hat
{p}\right)  ,\text{ \ }\hat{a}^{\dagger}=\frac{1}{\sqrt{2}}\left(  \frac
{\hat{x}}{l}-\frac{il}{\hbar}\hat{p}\right)  ,\nonumber\\
&  \hat{x}=l\frac{\hat{a}+\hat{a}^{\dagger}}{\sqrt{2}},\text{ \ }\hat{p}%
=\frac{\hbar}{l}\frac{\hat{a}-\hat{a}^{\dagger}}{i\sqrt{2}},\text{ \ }\left[
\hat{x},\hat{p}\right]  =i\hbar, \label{23}%
\end{align}
where $l$-parameter has the dimension of length. In terms of these operators,
(\ref{1}) reads,%
\begin{equation}
\hat{H}=\frac{\hat{p}^{2}}{2m}+\frac{k}{2}\hat{x}^{2}+\frac{\Omega}{2}\left(
\hat{x}\hat{p}+\hat{p}\hat{x}\right)  +F\hat{x}+V\hat{p}+\mathcal{E}.
\label{24}%
\end{equation}
The time-dependent quantities $m=m\left(  t\right)  $, $k=k\left(  t\right)
$, $\Omega=\Omega\left(  t\right)  ,$\ $F=F\left(  t\right)  $, $V=V\left(
t\right)  $ and $\mathcal{E}=\mathcal{E}\left(  t\right)  $ are related with
$\beta$, $\alpha$, $\gamma$ and $\delta$ in the form
\begin{align}
&  \frac{1}{m}=\frac{l^{2}}{\hbar}\operatorname{Re}\left(  \beta
-\alpha\right)  ,\text{ \ }k=\frac{\hbar}{l^{2}}\operatorname{Re}\left(
\beta+\alpha\right)  ,\text{ \ }\Omega=\operatorname{Im}\left(  \alpha\right)
,\text{ \ }\nonumber\\
&  F=\frac{\sqrt{2}\hbar}{l}\operatorname{Re}\left(  \gamma\right)  ,\text{
\ }V=\sqrt{2}l\operatorname{Im}\left(  \gamma\right)  ,\text{ \ }%
\mathcal{E}=\hbar\left(  \delta-\frac{\beta}{2}\right)  , \label{25}%
\end{align}
with the following initial conditions $m_{0}=m\left(  0\right)  $,
$k_{0}=k\left(  0\right)  $, $\Omega_{0}=\Omega\left(  0\right)  $,
$F_{0}=F\left(  0\right)  $, $V_{0}=V\left(  0\right)  $ and $\mathcal{E}%
_{0}=\mathcal{E}\left(  0\right)  $. Furthermore, it can be interesting to
find the inverse relation above, which is given by%
\begin{align}
\beta &  =\frac{l^{2}}{2\hbar}\left(  k+\frac{\hbar^{2}}{l^{4}m}\right)
,\text{ \ }\alpha=\frac{l^{2}}{2\hbar}\left(  k-\frac{\hbar^{2}}{l^{4}%
m}\right)  +i\Omega,\text{ \ }\nonumber\\
\gamma &  =\frac{l}{\hbar\sqrt{2}}\left(  F+\frac{i\hbar}{l^{2}}V\right)
,\text{ \ }\delta=\frac{1}{\hbar}\mathcal{E}+\frac{l^{2}}{4\hbar}\left(
k+\frac{\hbar^{2}}{l^{4}m}\right)  . \label{25a}%
\end{align}

Analyzing the relationships above can interpret the role of each parameter
present in the Hamiltonian (\ref{1}) and its effects on the squeezed and
displacement parameters, as will be seen later.

From (\ref{8}) and (\ref{23}), we can write $\hat{x}$ and $\hat{p}$ in terms
of $\hat{A}$ and $\hat{A}^{\dagger}$ in the form
\begin{align}
\hat{x}  &  =l\frac{\left(  f^{\ast}-g^{\ast}\right)  \hat{A}+\left(
f-g\right)  \hat{A}^{\dagger}}{\sqrt{2}}+\sqrt{2}l\operatorname{Re}u,\text{
\ }\nonumber\\
\hat{p}  &  =\frac{\hbar}{l}\frac{\left(  f^{\ast}+g^{\ast}\right)  \hat
{A}-\left(  f+g\right)  \hat{A}^{\dagger}}{i\sqrt{2}}+\sqrt{2}\frac{\hbar}%
{l}\operatorname{Im}u. \label{26}%
\end{align}

\subsection{Equations for $f$, $g$, and $\varphi$}

Substituting the representations (\ref{2}) and (\ref{3}) into (\ref{4}), we
may derive a set of differential equations for the functions $f$, $g$ and
$\varphi$,%
\begin{equation}
i\dot{f}=\alpha^{\ast}g-\beta f,\text{ \ }i\dot{g}=\beta g-\alpha f,\text{
\ }i\dot{\varphi}=\gamma^{\ast}g-\gamma f. \label{5}%
\end{equation}
Taking in account the initial conditions $f_{0}=f\left(  0\right)  =\left\vert
f_{0}\right\vert e^{i\theta_{1}}$ and $g_{0}=g\left(  0\right)  =\left\vert
g_{0}\right\vert e^{i\theta_{2}}$ we may obtain any nontrivial solution for
the functions $f$ and $g$. Then the function $\varphi$ can be found by a
simple integration%
\begin{equation}
\varphi=i\int_{0}^{t}\left(  \gamma f-\gamma^{\ast}g\right)  d\tau+\varphi
_{0}, \label{6}%
\end{equation}
where $\varphi_{0}=\varphi\left(  0\right)  =\left\vert \varphi_{0}\right\vert
e^{i\varsigma}$ is an arbitrary complex constant.

In order to obtain the solution for the functions $f$ and $g$, one may
identify the two first equations in (\ref{5}) with, for instance, the
well-known ones spin equation \cite{Bag2005},%
\begin{align}
&  i\dot{\psi}=\left(  \mathbf{K}\cdot\boldsymbol{\sigma}\right)
\psi,\nonumber\\
&  \psi=\left(
\begin{array}
[c]{c}%
f\\
g
\end{array}
\right)  ,\text{ \ }\mathbf{K}=-\left(  i\operatorname{Im}\alpha
,-i\operatorname{Re}\alpha,\beta\right)  ,\text{ \ }\boldsymbol{\sigma
}=\left(  \sigma_{1},\sigma_{2},\sigma_{3}\right)  , \label{7}%
\end{align}
where $\boldsymbol{\sigma}$ are the Pauli matrices.

For sake simplicity, if we assume that $\alpha$, $\beta$, and $\gamma$ are
time-independent, then we can write:%
\begin{equation}
\ddot{f}+\Theta^{2}f=0,\text{ \ }\Theta=\sqrt{\beta^{2}-\left\vert
\alpha\right\vert ^{2}}. \label{9}%
\end{equation}
This equation has the form of the simple harmonic oscillator equation with a
time-independent frequency $\Theta$. Given the exact solution of (\ref{9}),
one may obtain $g$ through the equation%
\begin{equation}
g=\frac{i}{\alpha^{\ast}}\dot{f}+\frac{\beta}{\alpha^{\ast}}f. \label{10}%
\end{equation}
Thus, the general solution of these equations is%
\begin{align}
&  f=f_{0}\cos\left(  \Theta t\right)  +i\left(  \beta f_{0}-\alpha^{\ast
}g_{0}\right)  \frac{\sin\left(  \Theta t\right)  }{\Theta},\text{ \ }%
g=g_{0}\cos\left(  \Theta t\right)  +i\left(  \alpha f_{0}-\beta g_{0}\right)
\frac{\sin\left(  \Theta t\right)  }{\Theta},\nonumber\\
&  \varphi=i\left(  \gamma f_{0}-\gamma^{\ast}g_{0}\right)  \frac{\sin\left(
\Theta t\right)  }{\Theta}+\left[  \left(  \gamma\beta-\gamma^{\ast}%
\alpha\right)  f_{0}+\left(  \gamma^{\ast}\beta-\gamma\alpha^{\ast}\right)
g_{0}\right]  \frac{\cos\left(  \Theta t\right)  -1}{\Theta^{2}}+\varphi_{0}.
\label{10a}%
\end{align}

\section{Time-dependent CSS}

Here, let us introduce a nonunitary exponential operator $\hat{S}$, as seen
below
\begin{equation}
\hat{S}=\exp\left(  \xi\hat{a}^{\dagger}+\frac{1}{2}\zeta\hat{a}^{\dagger
2}\right)  , \label{11}%
\end{equation}
where the time-dependent quantities $\xi=\xi\left(  t\right)  $ and
$\zeta=\zeta\left(  t\right)  $ corresponds to displacement and squeezed
parameters, respectively, in a nonunitary approach to CSS.

Taking into account that%
\begin{equation}
\xi=\frac{\varphi}{f},\text{ \ }\zeta=\frac{g}{f}, \label{11a}%
\end{equation}
and using the Baker--Campbell--Hausdorff theorem%
\begin{equation}
e^{A}Be^{-A}=B+\left[  A,B\right]  +\frac{1}{2}\left[  A,\left[  A,B\right]
\right]  +\ldots, \label{12}%
\end{equation}
one can express the canonical operator $\hat{a}$ through of the integral of
motion $\hat{A}$ as follows%
\begin{equation}
\hat{a}=\frac{1}{f}\hat{S}\hat{A}\hat{S}^{-1}. \label{13}%
\end{equation}

The application from this relation on the vacuum state $\left\vert
0\right\rangle $, which satisfies the annihilation condition $\hat
{a}\left\vert 0\right\rangle =0$, yields:%
\begin{equation}
\hat{A}\left\vert \xi,\zeta\right\rangle =0,\text{ } \tag{21a}\label{14}%
\end{equation}
where the most general state is given by%

\begin{equation}
\text{\ }\left\vert \xi,\zeta\right\rangle =\Phi\exp\left(  -\xi\hat
{a}^{\dagger}-\frac{1}{2}\zeta\hat{a}^{\dagger2}\right)  \left\vert
0\right\rangle . \tag{21b}\label{15}%
\end{equation}
From here, we call the states $\left\vert \xi,\zeta\right\rangle $ of
time-dependent CSS. Note that the function $\Phi=\Phi\left(  t\right)  $ was
introduced and will be following be determined in such a way that the states
$\left\vert \xi,\zeta\right\rangle $ satisfies the Schr\"{o}dinger's equation
(\ref{2}).

Substituting $\left\vert \xi,\zeta\right\rangle $ into (\ref{2}), we obtain
the following equation for $\Phi$:\setcounter{equation}{21}%
\begin{equation}
\frac{\dot{\Phi}}{\Phi}=\frac{\left\langle \zeta,\xi\left\vert \hat
{a}^{\dagger}\right\vert \xi,\zeta\right\rangle }{\left\langle \zeta,\xi
|\xi,\zeta\right\rangle }\dot{\xi}+\frac{1}{2}\frac{\left\langle \zeta
,\xi\left\vert \hat{a}^{\dagger2}\right\vert \xi,\zeta\right\rangle
}{\left\langle \zeta,\xi|\xi,\zeta\right\rangle }\dot{\zeta}-\frac{i}{\hbar
}\frac{\left\langle \zeta,\xi\left\vert \hat{H}\right\vert \xi,\zeta
\right\rangle }{\left\langle \zeta,\xi|\xi,\zeta\right\rangle }. \label{16}%
\end{equation}
Using the representation (\ref{8}) and the condition (\ref{14}), one can
calculate the mean values easily in (\ref{16}), as seen below%
\begin{align}
&  \frac{\left\langle \zeta,\xi\left\vert \hat{a}^{\dagger}\right\vert
\xi,\zeta\right\rangle }{\left\langle \zeta,\xi|\xi,\zeta\right\rangle
}=u^{\ast},\text{ \ }\frac{\left\langle \zeta,\xi\left\vert \hat{a}^{\dagger
2}\right\vert \xi,\zeta\right\rangle }{\left\langle \zeta,\xi|\xi
,\zeta\right\rangle }=u^{\ast2}-fg^{\ast},\nonumber\\
&  \frac{\left\langle \zeta,\xi\left\vert \hat{H}\right\vert \xi
,\zeta\right\rangle }{\left\langle \zeta,\xi|\xi,\zeta\right\rangle }%
=\hbar\operatorname{Re}\left[  2\gamma^{\ast}u+\alpha^{\ast}u^{2}-\alpha
fg^{\ast}+\beta\left(  \left\vert g\right\vert ^{2}+\left\vert u\right\vert
^{2}\right)  +\delta\right]  . \label{17}%
\end{align}
Thus, we can rewrite the $\Phi$ function, and so obtain
\begin{align}
&  \Phi=\frac{C}{\sqrt{f}}\exp\left(  \frac{g^{\ast}}{f}\frac{\varphi^{2}}%
{2}-\frac{\left\vert \varphi\right\vert ^{2}}{2}+i\phi\right)  ,\nonumber\\
&  \phi=\frac{1}{2}\int_{0}^{t}\left(  \beta-2\delta\right)  d\tau, \label{18}%
\end{align}
where $C$ is a normalization constant, which one may choose $C=1$ such that
$\left\langle \zeta,\xi|\xi,\zeta\right\rangle =1$.

Then, normalized time-dependent CSS that satisfies the Schr\"{o}dinger's
equation have the form%
\begin{equation}
\left\vert \xi,\zeta\right\rangle =\frac{1}{\sqrt{f}}\exp\left(  \frac
{g^{\ast}}{f}\frac{\varphi^{2}}{2}-\frac{\left\vert \varphi\right\vert ^{2}%
}{2}+i\phi\right)  \exp\left(  -\xi\hat{a}^{\dagger}-\frac{1}{2}\zeta\hat
{a}^{\dagger2}\right)  \left\vert 0\right\rangle . \label{18a}%
\end{equation}

\subsection{CSS in Fock-states representation}

The representation of the CSS via Fock-states can be obtained by taking into
account the generation function of the Hermite polynomials $H_{n}\left(
y\right)  $; see the formula (10.13.19) in \cite{Ede1953},%
\begin{equation}
\exp\left(  2yz-z^{2}\right)  =%
%TCIMACRO{\dsum \limits_{n=0}^{\infty}}%
%BeginExpansion
{\displaystyle\sum\limits_{n=0}^{\infty}}
%EndExpansion
\frac{H_{n}\left(  y\right)  }{n!}z^{n}, \label{19}%
\end{equation}
by applying it to the exponential operator function in (\ref{18a}),
\begin{equation}
\left\vert \xi,\zeta\right\rangle =\frac{1}{\sqrt{f}}\exp\left(  \frac
{g^{\ast}}{f}\frac{\varphi^{2}}{2}-\frac{\left\vert \varphi\right\vert ^{2}%
}{2}+i\phi\right)
%TCIMACRO{\dsum \limits_{n=0}^{\infty}}%
%BeginExpansion
{\displaystyle\sum\limits_{n=0}^{\infty}}
%EndExpansion
\left(  \frac{g}{2f}\right)  ^{\frac{n}{2}}H_{n}\left(  \frac{\varphi}%
{\sqrt{2gf}}\right)  \frac{\left(  -1\right)  ^{n}}{\sqrt{n!}}\left\vert
n\right\rangle . \label{20}%
\end{equation}

Here, it is important to highlight that the expansion of time-dependent CSS on
the time-independent Fock-states is obtained directly in the nonunitary
approach. On the other hand, following the usual construction of these states
through integrals of motion, it is unclear how it is possible to obtain an
equivalent expansion, see, e.g., \cite{Bag2015}. These states, Eq.\ (\ref{20}%
), allow us to obtain the usual CS and squeezed states (SS) classes, as seen below.

\begin{enumerate}
\item[1.] The particular case $\varphi=0\Rightarrow\xi=0$ leads to the
expression of the time-dependent SS%
\begin{equation}
\left\vert 0,\zeta\right\rangle =\left\vert \frac{g}{f}\right\rangle
=\frac{\exp\left(  i\phi\right)  }{\sqrt{f}}%
%TCIMACRO{\dsum \limits_{n=0}^{\infty}}%
%BeginExpansion
{\displaystyle\sum\limits_{n=0}^{\infty}}
%EndExpansion
\left(  -\frac{g}{f}\right)  ^{n}\frac{\sqrt{\left(  2n\right)  !}}{2^{n}%
n!}\left\vert 2n\right\rangle . \label{20a}%
\end{equation}
From (\ref{5}), we have that the state $\left\vert 0,\zeta\right\rangle $
satisfies the Schr\"{o}dinger's equation only if $\gamma=0$.

\item[2.] On the other hand, taking the condition $g=0\Rightarrow\zeta=0$
yields the expression of the time-dependent CS%
\begin{equation}
\left\vert \xi,0\right\rangle =\left\vert \frac{\varphi}{f}\right\rangle
=\frac{1}{\sqrt{f}}\exp\left(  i\phi-\frac{\left\vert \varphi\right\vert ^{2}%
}{2}\right)
%TCIMACRO{\dsum \limits_{n=0}^{\infty}}%
%BeginExpansion
{\displaystyle\sum\limits_{n=0}^{\infty}}
%EndExpansion
\left(  \frac{\varphi}{f}\right)  ^{n}\frac{\left(  -1\right)  ^{n}}{\sqrt
{n!}}\left\vert n\right\rangle . \label{20b}%
\end{equation}
From (\ref{5}) we have that the state $\left\vert \xi,0\right\rangle $
satisfies the Schr\"{o}dinger's equation only if $\alpha=0$.
\end{enumerate}

From the formula (10.13.22) in \cite{Ede1953}
\begin{equation}%
%TCIMACRO{\dsum \limits_{n=0}^{\infty}}%
%BeginExpansion
{\displaystyle\sum\limits_{n=0}^{\infty}}
%EndExpansion
H_{n}\left(  x\right)  H_{n}\left(  y\right)  \frac{z^{n}}{2^{n}n!}=\frac
{1}{\sqrt{1-z^{2}}}\exp\left[  \frac{2xyz-\left(  x^{2}+y^{2}\right)  z^{2}%
}{1-z^{2}}\right]  , \label{21}%
\end{equation}
one can easily see that the CSS are non-orthogonal to each other for arbitrary
values of the parameters $\xi$ and $\zeta$,%
\begin{align}
&  \left\langle \zeta_{1},\xi_{1}|\xi_{2},\zeta_{2}\right\rangle =\frac
{1}{\sqrt{f_{1}^{\ast}f_{2}-g_{1}^{\ast}g_{2}}}\exp\left(  \frac{\varphi
_{1}^{\ast}\varphi_{2}}{f_{1}^{\ast}f_{2}-g_{1}^{\ast}g_{2}}-\frac{\left\vert
\varphi_{1}\right\vert ^{2}+\left\vert \varphi_{2}\right\vert ^{2}}{2}\right)
\nonumber\\
&  \times\exp\left(  \frac{f_{1}^{\ast}g_{2}^{\ast}-f_{2}^{\ast}g_{1}^{\ast}%
}{f_{1}^{\ast}f_{2}-g_{1}^{\ast}g_{2}}\frac{\varphi_{2}^{2}}{2}-\frac
{f_{1}g_{2}-f_{2}g_{1}}{f_{1}^{\ast}f_{2}-g_{1}^{\ast}g_{2}}\frac{\varphi
_{1}^{\ast2}}{2}\right)  . \label{21a}%
\end{align}

Lastly, we can express the transition probability $P_{n}\left(  \xi
,\zeta\right)  =\left\vert \left\langle n|\xi,\zeta\right\rangle \right\vert
^{2}$ as follows%
\begin{equation}
P_{n}\left(  \xi,\zeta\right)  =\frac{1}{\left\vert f\right\vert }\exp\left[
\operatorname{Re}\left(  \frac{g^{\ast}\varphi^{2}}{f}\right)  -\left\vert
\varphi\right\vert ^{2}\right]  \left\vert H_{n}\left(  \frac{\varphi}%
{\sqrt{2gf}}\right)  \right\vert ^{2}\frac{1}{2^{n}n!}\left(  \frac{\left\vert
g\right\vert }{\left\vert f\right\vert }\right)  ^{n}, \label{22}%
\end{equation}
which coincides with the time-independent photon distribution function
\cite{Yue1976}.

\section{Semiclassical features}

In this section, we investigate the semiclassical features associated with CSS
by taking the mean value of some physical quantities and evaluate the
corresponding uncertainty relations.

\subsection{Mean values}

We begin investigating the mean values of the operators $\hat{x}$ and $\hat
{p}$ concerning the CSS. Taking into account the condition (\ref{14}) and the
representation (\ref{26}), we obtain that%

\begin{align}
&  \bar{x}=\bar{x}\left(  t\right)  =\left\langle \zeta,\xi\left\vert \hat
{x}\right\vert \xi,\zeta\right\rangle =\sqrt{2}l\operatorname{Re}u,\text{
\ }\bar{p}=\bar{p}\left(  t\right)  =\left\langle \zeta,\xi\left\vert \hat
{p}\right\vert \xi,\zeta\right\rangle =\sqrt{2}\frac{\hbar}{l}%
\operatorname{Im}u,\nonumber\\
&  \bar{x}_{0}=\bar{x}\left(  0\right)  =\sqrt{2}l\operatorname{Re}%
u_{0},\text{ \ }\bar{p}_{0}=\bar{p}\left(  0\right)  =\sqrt{2}\frac{\hbar}%
{l}\operatorname{Im}u_{0},\text{ \ }u_{0}=g_{0}\varphi_{0}^{\ast}-f_{0}^{\ast
}\varphi_{0}. \label{27}%
\end{align}
Here, one can see that there is a correspondence between the complex function
$\varphi$ and the mean values $\bar{x}$ and $\bar{p}$,%
\begin{equation}
\varphi=-\frac{1}{\sqrt{2}}\left(  \frac{f+g}{l}\bar{x}+il\frac{f-g}{\hbar
}\bar{p}\right)  . \label{28}%
\end{equation}

Using the Eq. (\ref{6}), one can verify that $\bar{x}$ and $\bar{p}$ evolves
following the Hamilton's equations,%
\begin{align}
&  \overset{\cdot}{\bar{x}}=\partial_{\bar{p}}H=\frac{\bar{p}}{m}+\Omega
\bar{x}+V,\text{ \ }\overset{\cdot}{\bar{p}}=-\partial_{\bar{x}}H=-k\bar
{x}-\Omega\bar{p}-F,\nonumber\\
&  H=\frac{\bar{p}^{2}}{2m}+\frac{k}{2}\bar{x}^{2}+\Omega\bar{x}\bar{p}%
+F\bar{x}+V\bar{p}+\mathcal{E}. \label{29}%
\end{align}
where $H$ is in the classical form of (\ref{24}).

\subsection{Standard deviation and uncertainty relations}

We recall that the standard deviation $\sigma_{\chi}=\sigma_{\chi}\left(
t\right)  $ and covariance $\sigma_{\chi\kappa}=\sigma_{\chi\kappa}\left(
t\right)  $ of certain physical quantities $\chi$ and $\kappa$ in some state
$\left\vert \psi\right\rangle $ is calculated via the corresponding operator
$\hat{\chi}$ and $\hat{\kappa}$, as follows:%
\begin{align}
&  \sigma_{\chi}\equiv\sqrt{\left\langle \left(  \hat{\chi}-\left\langle
\hat{\chi}\right\rangle \right)  ^{2}\right\rangle }=\sqrt{\bar{\chi}%
^{2}\left(  t\right)  -\left(  \bar{\chi}\left(  t\right)  \right)  ^{2}%
},\nonumber\\
&  \sigma_{\chi\kappa}\equiv\frac{\left\langle \left(  \hat{\chi}-\left\langle
\hat{\chi}\right\rangle \right)  \left(  \hat{\kappa}-\left\langle \hat
{\kappa}\right\rangle \right)  +\left(  \hat{\kappa}-\left\langle \hat{\kappa
}\right\rangle \right)  \left(  \hat{\chi}-\left\langle \hat{\chi
}\right\rangle \right)  \right\rangle }{2}=\frac{\overline{\chi\kappa}\left(
t\right)  +\overline{\kappa\chi}\left(  t\right)  }{2}-\bar{\chi}\left(
t\right)  \bar{\kappa}\left(  t\right)  ,\nonumber\\
&  \bar{\chi}^{2}\left(  t\right)  \equiv\left\langle \psi\left\vert \hat
{\chi}^{2}\right\vert \psi\right\rangle =\left\langle \hat{\chi}%
^{2}\right\rangle ,\text{ \ }\left(  \bar{\chi}\left(  t\right)  \right)
^{2}\equiv\left\langle \hat{\chi}\right\rangle ^{2},\text{ \ }\overline
{\chi\kappa}\left(  t\right)  \equiv\left\langle \hat{\chi}\hat{\kappa
}\right\rangle . \label{S.1}%
\end{align}

The standard deviation of the operators $\hat{x}$ and $\hat{p}$ and the
covariance $\sigma_{xp}$ with respect to the CSS is given by%
\begin{align}
&  \sigma_{x}=\sqrt{\bar{x}^{2}\left(  t\right)  -\left(  \bar{x}\left(
t\right)  \right)  ^{2}}=\frac{l}{\sqrt{2}}\left\vert f-g\right\vert ,\text{
\ }\sigma_{p}=\sqrt{\bar{p}^{2}\left(  t\right)  -\left(  \bar{p}\left(
t\right)  \right)  ^{2}}=\frac{\hbar}{l\sqrt{2}}\left\vert f+g\right\vert
,\nonumber\\
&  \sigma_{xp}=\overline{xp}\left(  t\right)  -\bar{x}\left(  t\right)
\bar{p}\left(  t\right)  -\frac{i\hbar}{2}=\hbar\operatorname{Im}\left(
fg^{\ast}\right)  . \label{S.2}%
\end{align}
From here, we can obtain the Heisenberg uncertainty relation%
\begin{equation}
\sigma_{x}\sigma_{p}=\frac{\hbar\left\vert f-g\right\vert \left\vert
f+g\right\vert }{2}=\frac{\hbar}{2}\sqrt{1+4\operatorname{Im}^{2}\left(
fg^{\ast}\right)  }\geq\frac{\hbar}{2},\text{ \ }\forall t. \label{S.4}%
\end{equation}

One can easily see that the condition $f=\mu g$, for $\mu$ real, minimizes the
Heisenberg uncertainty relation for any time \cite{Ped1987}. On the other
hand, it is seen from (\ref{5}) that the condition $f=\mu g$ necessarily leads
to $\alpha=0$. Note that $\alpha$ is the term that yields, under the condition
$\left\vert \alpha\right\vert \geq\beta$, a system with a continuous energy spectrum.

In the general case, the equality (\ref{S.4}) in $t=0$ is hold provided the
condition $\theta_{1}=\theta_{2}\equiv\theta$ is satisfied. Then, using the
relation $\left\vert g_{0}\right\vert ^{2}=\left\vert f_{0}\right\vert ^{2}-1$
together with the Eqs. (\ref{S.2}) and\ (\ref{S.4}), one can write $f_{0}$ and
$g_{0}$ in terms of the initial standard deviation $\sigma_{x_{0}}=\sigma
_{x}\left(  0\right)  $ as seen below%
\begin{align}
&  f_{0}=\frac{\sigma_{x_{0}}}{l\sqrt{2}}\left(  \frac{l^{2}}{2\sigma_{x_{0}%
}^{2}}+1\right)  e^{i\theta},\text{ \ }g_{0}=\frac{\sigma_{x_{0}}}{l\sqrt{2}%
}\left(  \frac{l^{2}}{2\sigma_{x_{0}}^{2}}-1\right)  e^{i\theta},\nonumber\\
&  \sigma_{p_{0}}=\frac{\hbar}{2\sigma_{x_{0}}}. \label{S.5}%
\end{align}
where we assume that $l\geq\sqrt{2}\sigma_{x_{0}}$.

In this case, we stress that the Schr\"{o}dinger-Robertson uncertainty
relation \cite{Rob1930} is minimized%
\begin{equation}
\sigma_{x}^{2}\sigma_{p}^{2}-\sigma_{xp}^{2}=\frac{\hbar^{2}}{4},\text{
\ }\forall t. \label{S.6}%
\end{equation}

\subsection{$x$-representation of the CSS}

In order to obtain the $x$-representation of the state (\ref{20}), as the
first step, we write the Fock-states as follows
\begin{equation}
\left\langle x|n\right\rangle =\Psi_{n}\left(  x\right)  =\frac{\left(
\hat{a}^{\dagger}\right)  ^{n}}{\sqrt{n!}}\Psi_{0}\left(  x\right)
=\frac{\left(  -1\right)  ^{n}}{\sqrt{2^{n}n!}}H_{n}\left(  \frac{x}%
{l}\right)  \Psi_{0}\left(  x\right)  , \label{32}%
\end{equation}
where the normalized state $\Psi_{0}\left(  x\right)  $ is calculated using
the annihilation condition $\hat{a}\Psi_{0}\left(  x\right)  =0$,
\begin{equation}
\Psi_{0}\left(  x\right)  =\frac{1}{\sqrt{l\sqrt{\pi}}}\exp\left(
-\frac{x^{2}}{2l^{2}}\right)  . \label{33}%
\end{equation}
Therefore, it follows from (\ref{20}), (\ref{21}) and (\ref{32}) the following
expression for the CSS in $x$-representation $\left\langle x|\xi
,\zeta\right\rangle =\Psi_{\xi,\zeta}\left(  x\right)  $,%
\begin{align}
\Psi_{\xi,\zeta}\left(  x\right)   &  =\frac{1}{\sqrt{f}}\exp\left(
\frac{g^{\ast}}{f}\frac{\varphi^{2}}{2}-\frac{\left\vert \varphi\right\vert
^{2}}{2}-\frac{i\vartheta}{\hbar}\right)  \Psi_{0}\left(  x\right)
%TCIMACRO{\dsum \limits_{n=0}^{\infty}}%
%BeginExpansion
{\displaystyle\sum\limits_{n=0}^{\infty}}
%EndExpansion
\left(  \frac{g}{f}\right)  ^{\frac{n}{2}}\frac{\left(  -1\right)  ^{n}}%
{2^{n}n!}H_{n}\left(  \frac{\varphi}{\sqrt{2gf}}\right)  H_{n}\left(  \frac
{x}{l}\right) \nonumber\\
&  =\frac{1}{\sqrt{l\sqrt{\pi}\left(  f-g\right)  }}\exp\left[  -\frac
{1}{2l^{2}}\frac{f+g}{f-g}\left(  x+\frac{\sqrt{2}l\varphi}{f+g}\right)
^{2}+\frac{f^{\ast}+g^{\ast}}{f+g}\frac{\varphi^{2}}{2}-\frac{\left\vert
\varphi\right\vert ^{2}}{2}-\frac{i\vartheta}{\hbar}\right] \nonumber\\
&  =\frac{1}{\sqrt{l\sqrt{\pi}\left(  f-g\right)  }}\exp\left[  -\frac
{f+g}{f-g}\frac{\left(  x-\bar{x}\right)  ^{2}}{2l^{2}}+\frac{i\bar{p}}%
{2\hbar}\left(  2x-\bar{x}\right)  -\frac{i\vartheta}{\hbar}\right]  ,
\label{36}%
\end{align}
where%
\begin{equation}
\vartheta=\int_{0}^{t}\left(  \mathcal{E}+\frac{V\bar{p}+F\bar{x}}{2}\right)
d\tau. \label{36a}%
\end{equation}

The corresponding probability densities take the form
\begin{equation}
\rho_{\xi,\zeta}\left(  t\right)  =\left\vert \Psi_{\xi,\zeta}\left(
x\right)  \right\vert ^{2}=\frac{1}{\sqrt{2\pi}\sigma_{x}}\exp\left[
-\frac{\left(  x-\bar{x}\right)  ^{2}}{2\sigma_{x}^{2}}\right]  . \label{37}%
\end{equation}

\section{Time-dependent harmonic oscillator}

We begin the discussion assuming a time-dependent dynamical system described
by the Hamiltonian%

\begin{equation}
\hat{H}=\frac{\hat{p}^{2}}{2m_{0}}+\frac{\varepsilon_{0}-\eta_{0}\cos\left(
\omega_{0}t\right)  }{2}\hat{x}^{2}. \label{39}%
\end{equation}
This equation is a particular case to the general quadratic
Hamiltonian\ (\ref{24}) with an oscillatory term that describes an external
force with a driving frequency $\omega_{0}$. Note that, if $\eta_{0}=0$, the
motion described by the Hamiltonian will be simple harmonic with resonant
frequency $\omega_{\varepsilon}=\sqrt{\varepsilon_{0}/m_{0}}$. The
Hamiltonian\ (\ref{24}) takes the form (\ref{39}) from considering the
following conditions:%

\begin{equation}
m=m_{0},\text{ \ }k=\varepsilon_{0}-\eta_{0}\cos\left(  \omega_{0}t\right)
,\text{ \ }\Omega=F=V=\mathcal{E}=0. \label{38}%
\end{equation}
From\ (\ref{25a}),\ (\ref{5}) and (\ref{38}), one can construct the integrals
of motion and find following equations%
\begin{align}
&  \dot{X}=\frac{il^{2}k}{\hbar}Y,\label{40a}\\
&  \dot{Y}=\frac{i\hbar}{l^{2}m_{0}}X, \label{40b}%
\end{align}
where we have used that $X=f+g$ and\ $Y=f-g$. In what follow, from\ these
expressions, one can write a second-order differential equation,
\begin{equation}
\ddot{Y}+\left[  \omega_{\varepsilon}^{2}-\omega_{\eta}^{2}\cos\left(
\omega_{0}t\right)  \right]  Y=0, \label{41}%
\end{equation}
with $\omega_{\varepsilon}=\sqrt{\frac{\varepsilon_{0}}{m_{0}}}\ $%
and\ $\omega_{\eta}=\sqrt{\frac{\eta_{0}}{m_{0}}}$. The Eq. (\ref{41})
describes the motion of a time-dependent harmonic oscillator which is
subjected to a driving force of form $\mathcal{F}\left(  t\right)
=\omega_{\eta}^{2}\cos\left(  \omega_{0}t\right)  $.

We can choose the rescaled \textquotedblleft time\textquotedblright\ variable
$\tau$ so that,%

\begin{equation}
\tau\left(  t\right)  =\frac{1}{2}\omega_{0}t,\text{ \ }\frac{d}{d\tau}%
=\frac{2}{\omega_{0}}\frac{d}{dt}.
\end{equation}

In this case, the Eq. (\ref{41}) can be identified with the Mathieu's equation
\cite{Mcl1964,Mathieu}:%
\begin{equation}
\ddot{Y}\left(  \tau\right)  +\left[  a-2q\cos\left(  2\tau\right)  \right]
Y\left(  \tau\right)  =0,
\end{equation}
where $a=4\omega_{\varepsilon}^{2}/\omega_{0}^{2}$ and $q=2\omega_{\eta}%
^{2}/\omega_{0}^{2}$.

Then, the general solution reads%
\begin{equation}
Y=c_{1}\mathrm{ce}_{\nu}\left(  \tau,q\right)  +c_{2}\mathrm{se}_{\nu}\left(
\tau,q\right)  , \label{43}%
\end{equation}
where $\mathrm{ce}_{\nu}\left(  \tau,q\right)  $ is cosine-elliptic, and
$\mathrm{se}_{\nu}\left(  \tau,q\right)  $ is sine-elliptic also referred to
as Mathieu's functions, given by \cite{Mcl1964},%

\begin{align}
\mathrm{ce}_{\nu}\left(  \tau,q\right)   &  =\cos\left(  \nu\tau\right)  +%
%TCIMACRO{\dsum \limits_{r=1}^{\infty}}%
%BeginExpansion
{\displaystyle\sum\limits_{r=1}^{\infty}}
%EndExpansion
q^{r}c_{r}\left(  \tau\right)  ,\nonumber\\
\mathrm{se}_{\nu}\left(  \tau,q\right)   &  =\sin\left(  \nu\tau\right)  +%
%TCIMACRO{\dsum \limits_{r=1}^{\infty}}%
%BeginExpansion
{\displaystyle\sum\limits_{r=1}^{\infty}}
%EndExpansion
q^{r}s_{r}\left(  \tau\right)  ,\nonumber\\
a  &  =\nu^{2}+%
%TCIMACRO{\dsum \limits_{r=1}^{\infty}}%
%BeginExpansion
{\displaystyle\sum\limits_{r=1}^{\infty}}
%EndExpansion
\mu_{r}q^{r}. \label{45}%
\end{align}
We should note that, taking into account the limit case $q=0$ $\left(
\omega_{\eta}=0\right)  $, the Eq. (\ref{41}) and its solutions reduce to
those of an unperturbed time-dependent harmonic oscillator.

It follows, substituting the solution (\ref{43})\ in Eq. (\ref{40b}), we
find,
\begin{equation}
X=\frac{l^{2}m_{0}\omega_{0}}{i\hbar}\left[  c_{1}\overset{\cdot}{\mathrm{ce}%
}_{\nu}\left(  \tau,q\right)  +c_{2}\overset{\cdot}{\mathrm{se}}_{\nu}\left(
\tau,q\right)  \right]  =-\frac{i}{2}\left[  c_{1}\overset{\cdot}{\mathrm{ce}%
}_{\nu}\left(  \tau,q\right)  +c_{2}\overset{\cdot}{\mathrm{se}}_{\nu}\left(
\tau,q\right)  \right]  , \label{46}%
\end{equation}
where\ $l^{2}=\hbar/m_{0}\omega_{0}$ and here, the dot above in $\mathrm{ce}%
_{\nu}\left(  \tau,q\right)  $ and $\mathrm{se}_{\nu}\left(  \tau,q\right)  $
denotes total derivative with respect to new variable $\tau$.

We may, in terms of the original functions $f$ and $g$, obtain the general
solution using the relations%
\begin{equation}
f=\frac{X+Y}{2},\text{ \ }g=\frac{X-Y}{2}. \label{48}%
\end{equation}
Then, the general solution has the form%

\begin{align}
f  &  =\frac{f_{0}}{2}\left[  \frac{2\mathrm{ce}_{\nu}\left(  \tau,q\right)
-i\overset{\cdot}{\mathrm{ce}}_{\nu}\left(  \tau,q\right)  }{2\mathrm{ce}%
_{\nu}\left(  0,q\right)  }-\frac{2\mathrm{se}_{\nu}\left(  \tau,q\right)
-i\overset{\cdot}{\mathrm{se}}_{\nu}\left(  \tau,q\right)  }{i\overset{\cdot
}{\mathrm{se}}_{\nu}\left(  0,q\right)  }\right] \nonumber\\
&  {}-\frac{g_{0}}{2}\left[  \frac{2\mathrm{ce}_{\nu}\left(  \tau,q\right)
-i\overset{\cdot}{\mathrm{ce}}_{\nu}\left(  \tau,q\right)  }{2\mathrm{ce}%
_{\nu}\left(  0,q\right)  }+\frac{2\mathrm{se}_{\nu}\left(  \tau,q\right)
-i\overset{\cdot}{\mathrm{se}}_{\nu}\left(  \tau,q\right)  }{i\overset{\cdot
}{\mathrm{se}}_{\nu}\left(  0,q\right)  }\right]  , \label{49a}%
\end{align}

\begin{align}
g  &  =\frac{g_{0}}{2}\left[  \frac{2\mathrm{ce}_{\nu}\left(  \tau,q\right)
+i\overset{\cdot}{\mathrm{ce}}_{\nu}\left(  \tau,q\right)  }{2\mathrm{ce}%
_{\nu}\left(  0,q\right)  }+\frac{2\mathrm{se}_{\nu}\left(  \tau,q\right)
+i\overset{\cdot}{\mathrm{se}}_{\nu}\left(  \tau,q\right)  }{i\overset{\cdot
}{\mathrm{se}}_{\nu}\left(  0,q\right)  }\right] \nonumber\\
&  -\frac{f_{0}}{2}\left[  \frac{2\mathrm{ce}_{\nu}\left(  \tau,q\right)
+i\overset{\cdot}{\mathrm{ce}}_{\nu}\left(  \tau,q\right)  }{2\mathrm{ce}%
_{\nu}\left(  0,q\right)  }-\frac{2\mathrm{se}_{\nu}\left(  \tau,q\right)
+i\overset{\cdot}{\mathrm{se}}_{\nu}\left(  \tau,q\right)  }{i\overset{\cdot
}{\mathrm{se}}_{\nu}\left(  0,q\right)  }\right]  . \label{49b}%
\end{align}

Hence, from these equations and Eqs. (\ref{27}), we derive the mean value of
the coordinate and momentum,%
\begin{align}
&  \bar{x}=\sqrt{\frac{2\hbar}{m_{0}\omega_{0}}}\operatorname{Re}\left(
g\varphi_{0}^{\ast}-f^{\ast}\varphi_{0}\right)  ,\text{ \ }\bar{p}%
=\sqrt{2\hbar m_{0}\omega_{0}}\operatorname{Im}\left(  g\varphi_{0}^{\ast
}-f^{\ast}\varphi_{0}\right)  ,\nonumber\\
&  \bar{x}_{0}=-2\left\vert \varphi_{0}\right\vert \sigma_{x_{0}}\cos\left(
\theta-\varsigma\right)  ,\text{ \ }\bar{p}_{0}=\frac{\left\vert \varphi
_{0}\right\vert \hbar}{\sigma_{x_{0}}}\sin\left(  \theta-\varsigma\right)
.\label{50}%
\end{align}
%

%TCIMACRO{\FRAME{ftFU}{3.0934in}{1.708in}{0pt}{\Qcb{Phase diagram for a
%one-dimensional time-dependent harmonic oscillator, with $f_{0}=1$ and
%$g_{0}=0$. The $q=0$ case (dashed curve) represents the classical solution,
%while the solid curve shows the phase-space trajectory to $q=1$.}}{\Qlb{fig1}%
%}{fig1.eps}{\special{ language "Scientific Word";  type "GRAPHIC";
%display "USEDEF";  valid_file "F";  width 3.0934in;  height 1.708in;
%depth 0pt;  original-width 16.9348in;  original-height 11.6949in;
%cropleft "0";  croptop "1";  cropright "1";  cropbottom "0";
%filename '../PLA_vfinal/fig1.eps';file-properties "XNPEU";}} }%
%BeginExpansion
\begin{figure}
[t]
\begin{center}
\includegraphics[
height=1.708in,
width=3.0934in
]%
{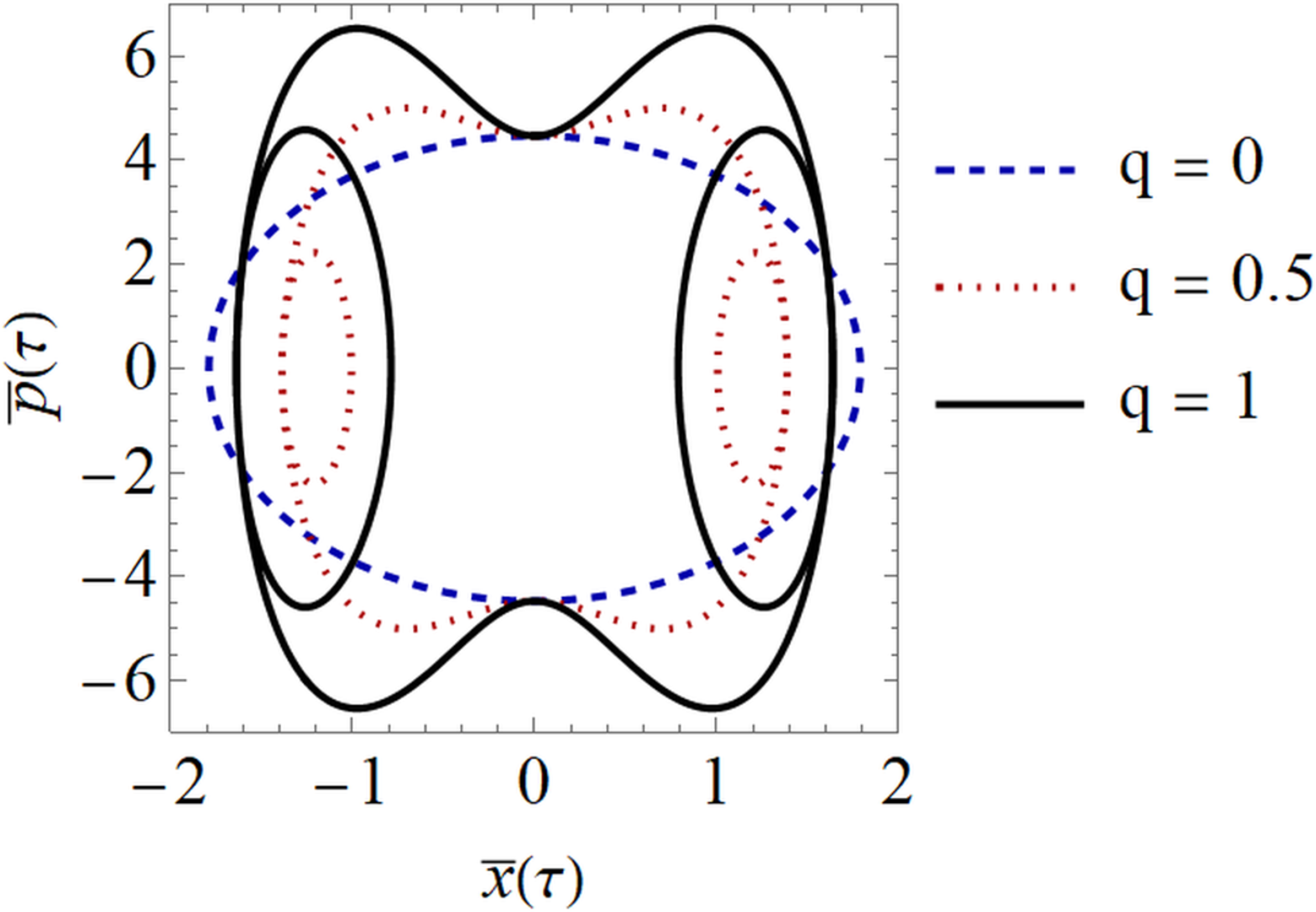}%
\caption{Phase diagram for a one-dimensional time-dependent harmonic
oscillator, with $f_{0}=1$ and $g_{0}=0$. The $q=0$ case (dashed curve)
represents the classical solution, while the solid curve shows the phase-space
trajectory to $q=1$.}%
\label{fig1}%
\end{center}
\end{figure}
%EndExpansion

The representation of the corresponding phase-space to this solution can be
view in Fig. \ref{fig1}. In this case, we can set the initial condition
$x\left(  0\right)  =0$ and $p\left(  0\right)  =4.5,$ and the other
parameters $\hbar=$ $m_{0}=\varepsilon_{0}=1$, $\omega_{0}=10$, $\eta_{0}=50$,
$\varphi_{0}=-i$, $\nu=1/2$ are chosen as an example.

Note that, for $q>0$ found similar orbits in phase-space of those ion
trajectories in Paul traps \cite{Trap}. However, we want to stress that the
limit case $q=0$ represented by the dashed curve, as shown in Fig. \ref{fig1},
reproduces the expected solution to simple harmonic motion.

\begin{figure}
[t]
\begin{center}
\includegraphics[
height=1.9873in,
width=2.9533in
]%
{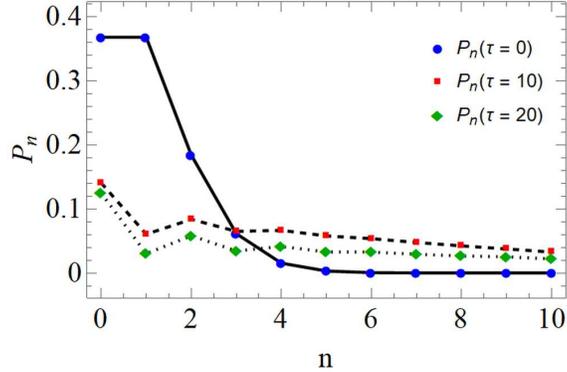}%
\caption{Transition probability distributions for the coherent squeezed states
of the time-dependent harmonic oscillator for $\tau=0$, $10$, and $20$.}%
\label{fig2}%
\end{center}
\end{figure}

Finally, we can now calculate the transition probability $P_{n}\left(
\xi,\zeta\right)  $ for the CSS by proceeding as discussed in previous
sections. In Fig. \ref{fig2}, considering different values of $\tau$ can be
viewed several transition probability from the Fock-states for the CSS, i.e.,
the probability to have $n$ number of excitations in the CSS.%

%TCIMACRO{\FRAME{ftFU}{2.9533in}{1.9873in}{0pt}{\Qcb{Transition probability
%distributions for the coherent squeezed states of the time-dependent harmonic
%oscillator for $\tau=0$, $10$, and $20$.}}{\Qlb{fig2}}{fig2.eps}%
%width 2.9533in;  height 1.9873in;  depth 0pt;  original-width 11.6949in;
%original-height 7.8369in;  cropleft "0";  croptop "1";  cropright "1";
%cropbottom "0";  filename '../PLA_vfinal/fig2.eps';file-properties "XNPEU";}}
%}%
%BeginExpansion
%EndExpansion

In this case, one can see that the distribution reached a maximum around the
value $n=0$, for all $\tau$, and the transition probability does not keep its
profile shape through time evolution.

\section{Final Remarks}

Given the recent interest the time-dependent systems in modern quantum
mechanics and their potential applications in many fields, we have
investigated the time-dependent displacement and squeezed parameters of the
CSS. Following the integral of motion method, we have constructed integrals of
motion for one-dimensional systems with general time-dependent quadratic
Hamiltonian. By applying a nonunitary transformation, we find a direct
relation between these integrals of motion $\hat{A}\left(  t\right)  $ and the
canonical annihilation operator $\hat{a}$. In this context, one has calculated
the corresponding CSS. As a consequence, we obtain the displacement
$\xi\left(  t\right)  $ and squeezed parameters $\zeta\left(  t\right)  $ in
terms of the functions $f\left(  t\right)  $, $g\left(  t\right)  $, and
$\phi\left(  t\right)  $. We have related these parameters with those physical
parameters present in the general quadratic Hamiltonian.

Moreover, by applying the nonunitary approach, one shown that CSS is
annihilated by the integral of motion. Additionally, the Schr\"{o}dinger
equation of the system to the case under consideration is satisfied. We have
also shown that the mean values of the phase-space variables, evaluate with
respect to time-dependent CSS, satisfy Hamilton's equations, and discussed
conditions under which the uncertainty relation is minimized.

We show that, considering the integrals of motion method in the nonunitary
approach, one can expand the time-dependent CSS on a time-independent Fock
basis. Thus, we can analyze systems with finite-levels, and, therefore, there
is potential for application in many fields of physics. This procedure ensures
the possibility of generations of these states even for physical systems
described by complex Hamiltonians.

Lastly, as an example, we described the solution to the general time-dependent
harmonic oscillator driven by a time-dependent driving force. In this case,
the phase diagram is presented and shown that, in limit case $q=0$, the simple
harmonic motion solution is recovered. Finally, one has obtained the
transition probability distributions that, as we showed, have not kept their
profile shape through time evolution.

\section{Acknowledgment}

ASP thanks the support of the Instituto Federal do Par\'{a}. The authors would
like to thank the referees for their valuable comments.

\end{document}